\documentclass[prx,twocolumn]{revtex4-2}
\usepackage{graphicx}
\usepackage{amsmath}
\usepackage{natbib}
\usepackage{epstopdf}
\usepackage{xcolor}
\newcommand{\bn}{\begin{eqnarray}}
\newcommand{\en}{\end{eqnarray}}
\newcommand{\eml}{\end{multline}}
\newcommand{\bml}{\begin{multline}}

\begin{document}

\title{General Machine Learning Algorithm for Quantum Teleportation}
 \author{Allison Brattley,$^{1,2}$ Tom\'a\v{s} Opatrn\'y,$^3$ and Kunal K. Das$^{4,5}$}
  \affiliation{$^1$Department of Physics, Yale University, New Haven, CT 06520, USA}
  \affiliation{$^2$ Department of Physics, Massachusetts Institute of Technology, Cambridge, MA 02139, USA}
  \affiliation{$^3$Department of Optics, Palacký University, 771 46 Olomouc, Czech Republic}
 \affiliation{$^4$Department of Physical Sciences, Kutztown University of Pennsylvania, Kutztown, Pennsylvania 19530, USA}
  \affiliation{$^5$Department of Physics and Astronomy, Stony Brook University, New York 11794-3800, USA}

\begin{abstract}
We present a general algorithm, based on machine learning, which can create optimal unitary operators to implement quantum teleportation in any system with well-defined set of measurements in a relevant entangled basis. We illustrate it with a collective spin model and demonstrate its versatility by applying it to teloportation of single and multiple qubit states, coherent and Dicke states, and for systems with prior distributions and unequal dimensions. All cases display significant regimes of quantum advantage over corresponding classical schemes with no entanglement. The algorithm offers the flexibility to choose a balance between target fidelity and computational cost.
\end{abstract}

\maketitle

\section{Introduction}

Since its conception \cite{Wootters1993} and first realization \cite{bouwmeester_experimental_1997-1}, quantum teleportation continues to be rapidly developed as a signature quantum technology on a variety of platforms, including, photons \cite{Polzik1998,marcikic_long-distance_2003}, atoms \cite{Jian-Wei_Pan-PNAS,Monroe_2009}, ions \cite{barrett_deterministic_2004,riebe_deterministic_2004}, solid state \cite{pfaff_unconditional_2014,steffen_deterministic_2013} and propagating microwaves \cite{fedorov_experimental_nodate}. Quantum teleportation has bridged ground and satellite \cite{ren_ground--satellite_2017}, spanned different platforms \cite{fiaschi_optomechanical_2021}, distant nodes in a network \cite{hermans_qubit_2022}, and has been applied to higher dimension states \cite{Pan2019} and to quantum gates \cite{wan_quantum_2019,feng_chip--chip_2025}.

The central feature of quantum teleportation \cite{Zwiebach} is the correlation that is established between, the specific measurement in an entangled basis and the resulting unitary transformation executed to remotely replicate the original input state.  In the original proposal with spin singlets, the measurement is done in a four-state Bell basis \cite{Wootters1993, Bell}. In contrast to this discrete variable teleportation, there are well-defined analytic schemes for the other limit of infinite particles, for continuous variable teleportation using measurements on conjugate variables like position and momentum distributions \cite{Polzik1998,Vaidman1994,BraunsteinKimble}. However, in general there is a vast range in between, of systems comprising of a finite number of particles or states, that are not amenable to such customized schemes that apply to the limiting cases. Furthermore, the idealized unitary operators and perfectly entangled states, often assumed in theory, cannot be realized in experiments.

Thus, it is desirable to have an efficient method to establish an optimal correlation between the space of all possible measurements in the entangled basis and the space of the resultant unitary transformations, even when conditions are less than ideal; and yet be sufficiently general that it can be applied to models that can be realized in a broad range of physical systems in experiments.  This paper provides an algorithm to do this using machine learning methods \cite{watkins_technical_1992}. The algorithm provides robust optimization under a wide variety of conditions, including having limited set of accessible unitary operations, nonuniform distribution of input states, imperfect entanglement, diverse input states and non-commensurate system sizes.

In Sec. II, we provide a detailed description of the general algorithm. We present a flexible physical model in Sec. III to demonstrate the utility of our algorithm. The versatility of the method is underscored in Sec. IV by applying it to the discrete variable Bell basis teleportation. In Sec. V, we apply our algorithm to N-particle spin coherent states. The strengths of the algorithm become apparent in Sec. VI, where we assume a nontrivial prior distribution of input states, and Sec. VII, where we teleport states between systems of unequal particle numbers. In Sec.~VIII, we demonstrate how our algorithm may be extended to include diverse states by applying it to teleport rotated Dicke states, and we conclude with a summary of our results and outlook in Sec.~IX. Two appendices contain derivations of some results used in the main body of the paper, as well as the details of our numerical simulations.

\section{Algorithm Description}

The algorithm comprises of the following steps:

1. Initiating systems A and B in some well-defined states, we entangle them by evolving via an interaction Hamiltonian $\hat{H}_{AB}$.  In the discrete limit this implies maximizing the entanglement entropy
$S(\rho_I) = -{\rm Tr}[\rho_I\ln(\rho_I)]$
with $\rho_{I}$ the reduced density matrix for one of $I=A,B$. In general, constrained by the kind of the measurement of the $AC$ system, maximum fidelity of teleportation is not assured by maximum entanglement. So we implement a machine learning algorithm to carry through the entire teleportation scheme as outlined below to optimize the fidelity. After varying multiple parameters, we found that the process can be streamlined by simply varying the interaction time. The evolution times for minimum variance of the measurement variables, and for maximum entropy do not generally coincide. We choose to time evolve to some intermediate time to optimize the tradeoff between entanglement and accurate measurement.

2. A and B are separated spatially, and  A is allowed to interact with system C prepared in a specific state $\psi_i$ sampled from a space $\{\psi_i\}$ of states that we would like to teleport. AC evolves via an interaction Hamiltonian to maximum achievable entanglement. In the case of a single qubit, the AC system is simply measured. For higher dimensional cases, states of A and C must be sufficiently entangled for the commuting variables to be measured  (counterparts of $x_+, p_-$ for EPR states).

3. The commuting observables, one each for A and C, are measured in a common basis $\Phi_{j}^{AC}$. Each combination of input state and measurement outcome corresponds to a specific output state of B, which we label $\psi_B$.

4. A unitary operation is done on B to recreate the input state of C. The optimal operation $\hat{U}_{ij}(\psi_{i},\Phi_{j}^{AC})$ is found by optimising some measure of fidelity with respect to a relevant set of parameters.

5. Steps 3-4 are iterated for every projection $\Phi_{j}^{AC}$ in the basis, creating a library of optimal unitary operations for each projection for a specific input state of C.

6. Steps 2-5 are iterated for sufficient sampling of input states from the set $\{\psi_{i}\}$.

7. Allowing the option of a prior distribution $p(\psi_{i})$, we integrate out the dependence on the input state to define the desired set of unitary operators $\hat{\mathbf{U}}_j(\Phi_{j}^{AC})$, one for each measurement outcome $\Phi_{j}^{AC}$. The aim is to achieve the maximum possible mean fidelity.

\begin{figure}[t]
\centering
%\vspace{-0.2\linewidth}
\includegraphics[width=\columnwidth]{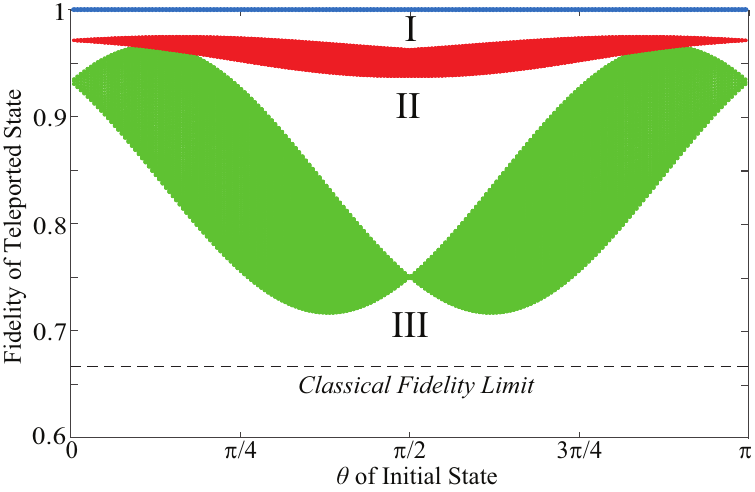}
\caption{Teleportation of single qubits, each point marking a specific input state. The four unitary operators are determined by optimizing: $I$. the three Euler angles with optimization of initial conditions; $II$. the three Euler angles without optimizing initial conditions; $III$. only $J_x$ and $J_y$ rotations but with partial optimization of initial conditions. The maximum classical fidelity (dashed line) is surpassed by all, with average fidelities, $I: 100\%, II: 95.7\%$ and $III: 81.1\%$.} \vspace{-5mm}
\label{bell_teleport}
\end{figure}

\section{Physical Model}

For a concrete implementation of the algorithm, we adopt a collective spin model that will allow us to explore versatile scenarios.  We assume two species of particles labelled $i=1,2$ described by bosonic operators, $\hat{a}_i$ and $\hat{b}_i$ that represent two discrete states for each species. We can define collective spin operators $\hat{J}_{xi}=\frac{1}{2}(\hat{a}_i^\dagger\hat{b}_i + \hat{a}_i\hat{b}_i^\dagger), \hat{J}_{yi}=\frac{1}{2i}(\hat{a}_i^\dagger\hat{b}_i - \hat{a}_i\hat{b}_i^\dagger),
\hat{J}_{zi}=\frac{1}{2}(\hat{a}_i^\dagger\hat{a}_i + \hat{b}_i^\dagger\hat{b}_i)$.
This is well suited for bridging the two limits: Two entangled particles yield single qubit teleportation, and for large particle number, assuming $\langle \hat{J}_x^2\rangle, \langle \hat{J}_y^2\rangle  \ll \langle \hat{J}_z^2\rangle\simeq N^2/4$ we can set $[\hat{J}_x, \hat{J}_y] =i\hat{J}_z \simeq \pm iN/2$ to have continuous variable EPR like states \cite{Vaidman1994,BraunsteinKimble}.

The state of system $C$ is to be teleported via the entanglement of spatially separated systems $A$ and $B$. Since $A$ and $C$ interact as well while $B$ and $C$ never do, we take $B$ and $C$ to be of the same species, so the interaction Hamiltonians $\hat{H}_{AB}$ and $\hat{H}_{AC}$ have the same form. A simple Hamiltonian, realizable with ultracold atoms \cite{Opatrny-Das} that serves our purpose is $\hat{H}=\hat{H}_L+\hat{H}_Q$:
\begin{eqnarray}
   \hat{H}_L &=& {\textstyle \sum_{i=1,2}}(\kappa_{xi}\hat{J}_{xi}+\kappa_{yi}\hat{J}_{yi}+\kappa_{zi}\hat{J}_{zi}) \nonumber \\
    \hat{H}_Q &=& \chi(\hat{J}_{x1}+\hat{J}_{x2})^2 + \chi (\hat{J}_{y1}+\hat{J}_{y2})^2,
    \label{HQ_def}
\end{eqnarray}
In our simulations our states $A,B$ are initiated as coherent states. The state to be teleported, $C$ is initiated in an eigenstate of the linear part $\hat{H}_L$ with two natural choices: Dicke states which are the eigenstates  $H_L|j,m\rangle=\hbar m|j,m\rangle$; and coherent states (eigenstates with $m=\pm j$) labeled by mean co-ordinates on the Bloch sphere, $|\theta, \phi\rangle$, equivalent to the rotated vacuum. We will consider both, but focus primarily on coherent states which are better accessed in experiments.

The quadratic part $H_Q$ in Eq.~(\ref{HQ_def}) has been shown to operate as components of $SU(1,1)$ and $SU(2)$ interferometers \cite{Opatrny-Das}. Generating entanglement via $SU(1,1)$ interaction implies the two coherent states are initialized on opposite poles of the Bloch sphere, and $SU(2)$ near the same pole. Maximal entanglement is easier to achieve starting with the states at opposite poles, hence $H_Q\rightarrow H_{AB}$ will operate as $SU(1,1)$. But for $H_Q\rightarrow H_{AC}$, to realize projections on entangled bases, we consider both choices.

\section{Single Qubit Teleportation}

We benchmark our algorithm by first applying it to single-qubit teleportation. In the canonical approach, the AB state is a Bell state, and the AC measurement fixes the unitary operations as Pauli matrices. We generate AB by evolving to the maximal entropy of the system, and follow our algorithm. In step 3, we directly measure AC in the Bell basis, $\Phi_j$ with $j=1,2,3,4$, to compare with the canonical method. The state of C is sampled from a uniform distribution of rotated qubit states covering the entire Bloch sphere. The unitary operations for the four possible measurements for each input state  $\psi_{i}$, are determined by maximizing its fidelity
\begin{eqnarray}\label{fidelity}
F_{ij} = |\langle \psi_i| \hat{U}_{ij} |\psi_B\rangle|^2.
\end{eqnarray}
with respect to the three Euler angles
\bn \hat{U}_{ij}(\psi_{i},\Phi_{j}^{AC})=e^{-i\gamma_{ij} \hat{J}_z}e^{-i\beta_{ij} \hat{J}_x}e^{-i\alpha_{ij} \hat{J}_z}.\en
We then average over all the input states by $\alpha_j={\rm arg}\left[\sum_{i}p(\psi_i){\rm exp}(i\alpha_{ij})\right]$ and likewise for the other Euler angles; here, $p(\psi_i)$ is a constant. This yields the set of four unitary operators, one for each measurement outcome
$\hat{\mathbf{U}}_j(\Phi_j^{AC}) = e^{-i\gamma_j \hat{J}_z}e^{-i\beta_j \hat{J}_x}e^{-i\alpha_j \hat{J}_z}$.

In Fig.~\ref{bell_teleport}, we compare the tradeoff between the fidelity and the number of parameters used for optimization; this becomes significant for larger particle numbers which require more computational resources. All our cases substantially surpass the classical teleporation fidelity limit of $2/3$ for all input states sampled. Case I shows that perfect fidelity is obtained when we optimize all the Euler angles to find the unitary operators, in tandem with choosing a favorable starting point for the gradient descent search for the best angles. Case II shows that fidelity decreases if we always use a random starting point. There are infinite choices of unitary operators for each input state to reach unit fidelity.  Averaging over the specific operators we find for each, is not optimal for all, and hence the vertical spread and less than unit fidelity. Case III shows a simplified and faster optimization where we use only a two-parameter rotation $\hat{\mathbf{U}}_j(\Phi_j^{AC}) = e^{-i(\theta^x_{j} \hat{J}_x+\theta^y_{j}\hat{J}_y)}$ about the $x$ and $y$ axes to determine the optimal unitary operators. This works better near the poles of the Bloch sphere, since near $J_z=0$, a rotation around the $z$ axis would be ideal. The vertical spread is due to the optimal unitary operators being better for some input states than others.

\begin{figure}[t]
\centering
%\vspace{-0.2\linewidth}
\includegraphics[width=\columnwidth]{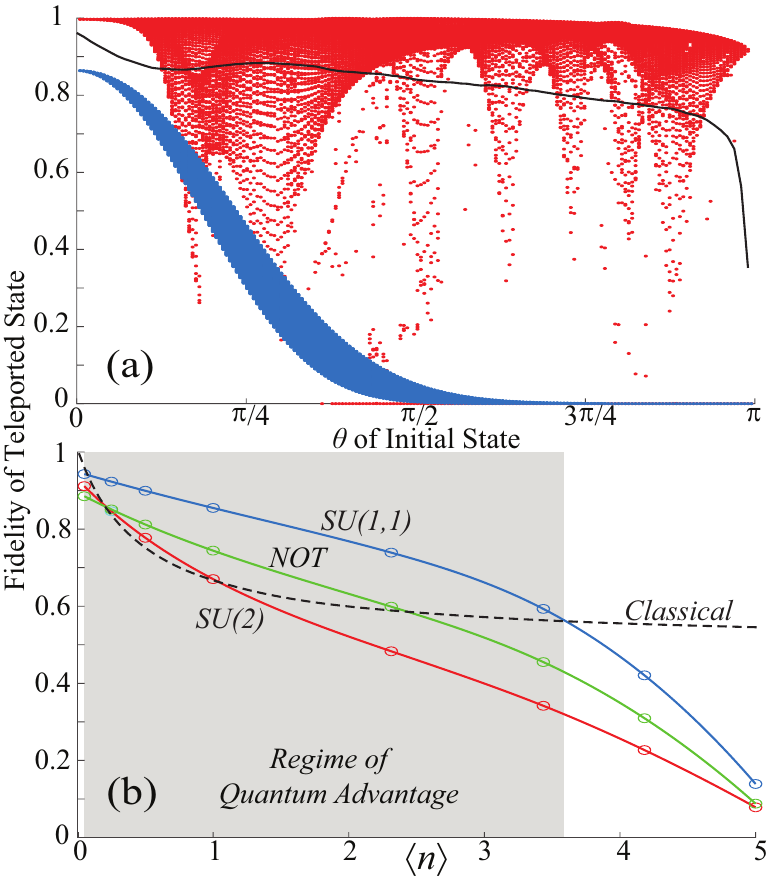}
\caption{(a) Fidelity of coherent states in SU(1,1) initialization for $10$ particles in A,B,C, for uniform distribution of input state of C. Red dots represent the optimized fidelity after applying $\hat{U}_{ij}(\psi_i,\Phi_j^{AC})$; the black line is their average; the blue band comprises of points marking the fidelity on applying $\hat{\mathbf{U}}_j(\Phi_j^{AC})$. (b) Averaged fidelities as a function of the occupation numbers $\langle n\rangle$ for a von Mises-Fisher prior distribution shown for SU(2), SU(1,1) initializations of C, and for teloporting $\bar{C}$. Comparison with the classical limit shown with as dashed line shows a broad regime of quantum advantage. }\vspace{-5mm}
\label{fidelity_figs}
\end{figure}

\section{N-particle spin coherent states}

We next consider $N$ particles in a spin coherent state.  We initiate A and B in an $SU(1,1)$ configuration, and take $J_x$ and $J_y$ as our measurement variables. If we time evolve this configuration by $H_{AB}$ with $\chi=1$ the upper and lower boundaries for the variances at any instant of time are given by
\begin{eqnarray}
    \sigma_{\pm}^2 = \langle J_{xA}^2 \rangle + \langle J_{xB}^2 \rangle \pm 2\sqrt{\langle J_{xA}J_{xB} \rangle^2 + \langle J_{xA}J_{yB} \rangle^2}
\end{eqnarray}
with all expectations evaluated at the same instant. The details of this derivation are provided in Appendix \ref{App:Var}. The state of C to be teleported is sampled as described for the single-quibit case, and we apply the interaction Hamiltonian to AB until an optimal time, with exact numerical values provided in Appendix \ref{App:Details}.  Once systems $A$ and $B$ are separated, we allow $A$ to interact with $C$ via interaction Hamiltonian $\hat{H}_{AC}\equiv -\hat{H}_{AB}$, the sign reversed to retrace the path traversed by the AB evolution.

Figure~\ref{fidelity_figs}(a) shows the outcome of our algorithm for $N_i=N=10$ particles, $i=A,B, C$. We use two parameter rotation as in case III for single qubit. After running steps 2 through 6, applying the optimal unitary operations
$\hat{U}_{ij}(\psi_{i},\Phi_{j}^{AC})=e^{-i(\theta^x_{ij} \hat{J}_x+\theta^y_{ij}\hat{J}_y)}$, the fidelity obtained are shown as a spread of red dots, one for each combination of input state $\psi_i$ and measurement of $AC$. Here fidelity is optimized assuming states B and C are localized near the same pole, a $SU(1,1)$ configuration. The spread of the dots is due to imperfect convergence due to limitations of the gradient descent method. The dot density cannot be discerned, so we also plot the average as a black line; the majority (about $81\%$) of outcomes are above $80\%$ fidelity.  After implementing step 7, by integrating over all initial states we determine $(N+1)^2$ unitary operations, one for each measurement outcome of $AC$, and apply those to the set of \emph{all} states to be teleported. The final fidelities lie in a narrow blue band in Fig.~\ref{fidelity_figs}(a). They are skewed towards one  pole, since the use of only two parameters favors polar regions where commutators of ${\hat J}_i$ are approximately constant, and the pole favored is set by the choice of $SU(1,1)$ or $SU(2)$.

\section{Prior Distribution}

The fidelity clearly has a strong dependence on the state teleported. This suggests that drawing states from a prior disribution can lead to substantial increase in mean fidelity. Analytical solutions for the classical benchmark that assume an uniform distribution of states to be teleported \cite{Massar-Popescu-PRL1995} cannot be applied when states have a non-uniform prior distribution \cite{Braunstein2000,Polzik2006,Cirac2005}. But our algorithm can be applied to states with prior distribution with equal facility. We illustrate with a von Mises-Fisher distribution
$ {\cal P}(\theta, \phi)=\beta e^{\beta \cos(\theta)}/[4\pi \sinh(\beta)]$,
the analog of a thermal distribution on a sphere; $\pm|\beta|$ specifies the pole where the distribution peaks. Each coherent state $|\theta, \phi\rangle$ has mean occupation $\bar{n}=N\sin^2(\theta/2)$, so averaging over the Bloch sphere with weights ${\cal P}(\theta, \phi)$ we get  $\langle n\rangle =(N/2)(1-\coth(\beta)+\beta^{-1})$.

In our algorithm, we now set $p(\psi_i)\equiv {\cal P}(\theta, \phi)$ to compute the averages in step 7. We tested our algorithm for different prior distributions by varying $\beta$, and in Fig.~\ref{fidelity_figs}(b), we plot the mean fidelity of teleportation as a function of the expected occupation number $\langle n\rangle$ for that distribution. Both $SU(2)$ and $SU(1,1)$ interactions between $A$ and $C$ are considered, with the latter displaying better fidelity. For $SU(2)$, evolving the $AC$ system via $H_{AC}\equiv H_{AB}$, (no sign reversal) for duration $\pi/(4N_A)$ leads to maximum entanglement \cite{Opatrny-Das}; and the optimization is done after rotating the output state $B$ by $\pi$ around the $J_x$ axis. As this axis is not always optimal, the $SU(2)$ fidelities are lower, and could be further improved. The fidelity approaches unity with decreasing $\langle n\rangle$ as the  distribution becomes increasingly peaked at one pole.

We compare our fidelity outcomes with the maximum classical fidelity that is achievable with no entanglement. We obtain this by considering the limit of infinite number of particles which marks the other extreme of continuous variable teleportation. With no entanglement, the fidelity can be computed analytically for coherent states that obey a thermal distribution \cite{Braunstein2000}; the maximum average fidelity possible is
\bn F_{cl} = \frac{\langle n \rangle +1}{2\langle n \rangle +1 },\en where $\langle n \rangle$ is the mean number of excitations of the thermal ensemble of input states. We have shown that this expression provides an upper bound to the maximum fidelity for classical teleportation for finite particle number \cite{Paper2}; so we use this to benchmark the fidelity of teleportation. The $F_{cl}$ values are plotted as a dashed line in Figure~\ref{fidelity_figs}(b). We observe a significant regime of quantum advantage in teleportation fidelity achieved using our algorithm.

Figure~\ref{fidelity_figs}(b) also shows fidelities for teleporting a NOT state, where we interact systems $A,C$ in a $SU(2)$ configuration and teleport the state $\bar{C}$ antipodal to it. Comparable fidelities are achieved across all prior distributions. Due to the mean fidelity being less than one and larger number of particles used, this does not violate the impossibility of creating a universal NOT gate \cite{Fiurasek_2001,Mista-Fiurasek_2006}.

\begin{figure}[t]
\centering
%\vspace{-0.2\linewidth}
\includegraphics[width=\columnwidth]{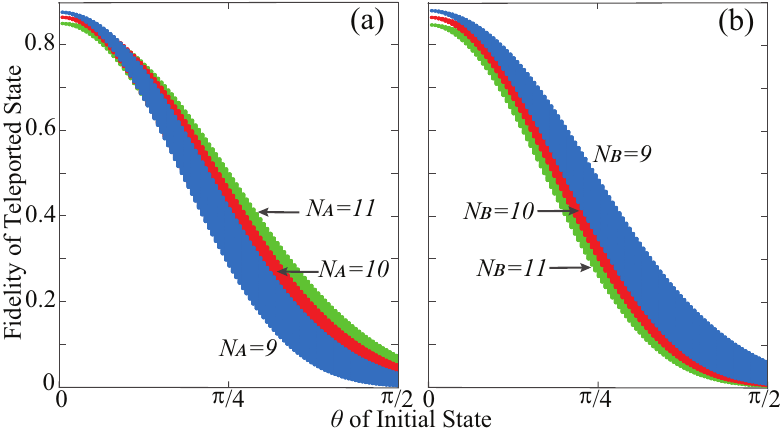}
\caption{The fidelities when the number of atoms among the species are different. (a) $N_B = N_C = 10$, and $N_A = 9$, $10$, and $11$. (b) $N_A = N_C = 10$, and $N_B = 9$, $10$, and $11$.}\vspace{-5mm}
\label{unequal_number}
\end{figure}

\section{Unequal Numbers of Particles}

To allow for possible number fluctuations, we now show that the same teleportation scheme may be implemented if A,B,C have unequal numbers of particles. In creating the unitary operators, we assume equal number of particles in all. But in implementing teleporatation for an arbitrary initial state, we allow the systems to have unequal numbers. The measurements are now communicated differently:  States with variable numbers of particles may be visualized as states on Bloch spheres with radii that increase with particle number. For coherent states, measurements may be communicated by sharing the coordinates $(r,\theta,\phi)$ on the sphere, keeping the same angles $\theta, \phi$, and changing the radius to match that of the Bloch sphere of the relevant sub-group. In practice, we compute the fidelity in Eq.~(\ref{fidelity}) with $\psi_C\rightarrow \tilde{\psi}_C$ where they share the $\kappa_{ji}$ values to specify orientation on the Bloch sphere but the eigenstates of $\hat{J}_{x,y,z}$ used in $\tilde{\psi}_C$ belong to the Hilbert space of $B$.

%In Fig.~\ref{unequal_number} (a) we compare the fidelities of teleportation for A having one more or less particle than B and C; and in Fig.~\ref{unequal_number} (b) we show the case of B being similarly varied with A and C both having 10 particles.
In Fig.~\ref{unequal_number}(a) we compare the fidelities of teleportation $N_B=N_C=10=N_A\pm 1$; and in Fig.~\ref{unequal_number}(b) we show the case $N_A=N_C=10=N_B\pm 1$.
We can see the fidelity is quite robust against about $10\%$ fluctuations of the pair of entangled states. Fig.~\ref{unequal_number}(b) shows a widening of the fidelity with reduced particle number, due to the general widening of the spread of the states.

\begin{figure}[t]
\centering
%\vspace{-0.2\linewidth}
\includegraphics[width=\columnwidth]{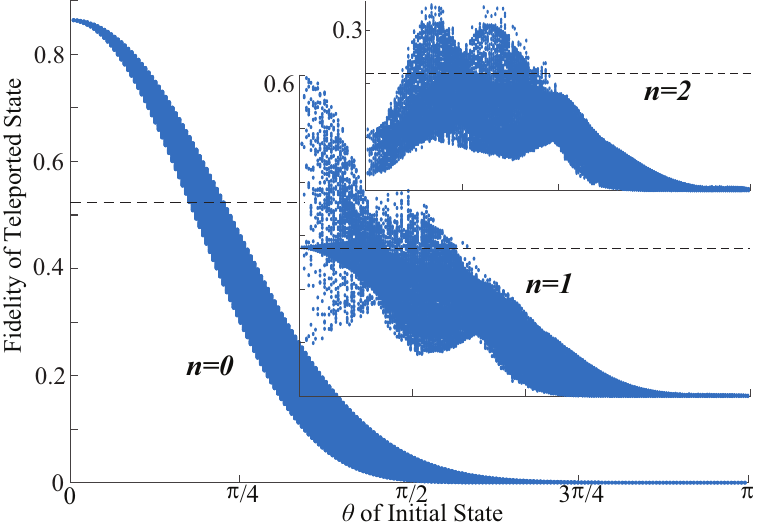}
\caption{The fidelities of  the first three Dicke states are plotted. All the states have regimes where a significant fraction of the states surpass the maximum fidelity without entanglement, shown as dashed lines.}\vspace{-5mm}
\label{Dicke}
\end{figure}

\section{Rotated Dicke States}

As an extension and validation for our scheme, we use the libraries built with coherent states, and apply it to teleport Dicke states arbitrarily rotated on the Bloch sphere. To derive a classical comparison in the limit of no entanglement, let us assume that an analogue to heterodyne detection is performed on an $N$-qubit system, in which the state is projected on an overcomplete set of spin states. The conditional probability of obtaining a particular coherent state $|\theta,\phi\rangle$, assuming that the system was in the Dicke state $|n\rangle$, is
\begin{eqnarray}
  \mathcal{P}(\theta,\phi|n) &=& \frac{(N+1)!}{4\pi n! (N-n)!}\cos^{2n}\frac{\theta}{2}\sin^{2(N-n)}\frac{\theta}{2},
\end{eqnarray}
where $\int_{0}^{2\pi}\int_{0}^{\pi}\mathcal{P}(\theta,\phi|n)\sin\theta d\theta d\phi = 1$. If the state $|\theta,\phi\rangle$ is detected, and the same state $|\theta,\phi\rangle$ is created, the fidelity with respect to the original state $|n\rangle$ is equal to $|\langle \theta,\phi|n\rangle|^2$, such that the mean fidelity is \cite{Paper2}
\begin{eqnarray}
  \bar{F}_n &=& \frac{(N+1)!N!(2n)!(2N-2n)!}{[n!(N-n)!]^2(2N+1)!}.
\end{eqnarray}

The fidelities for the lowest three states are shown in Fig.~\ref{Dicke}, compared to the maximum fidelity achievable without entanglement shown as dashed lines. All the states show a significant regime where the fidelities are above that limit. Specifically, for $n=0,1$ there is a non-trivial regime near one poles where the average fidelity is decidedly above the non-entangled limit. In the range where some of the fidelity exceeds the non-entangled limit, the three states have $80.2\%,16.8\%$ and $10.6\%$ above the dashed line. Lower fidelities are to be expected with Dicke states since they are circular bands on the Bloch sphere, so simple rotations will not suffice to overlap two Dicke states having a completely different mean radii. But the fact that even just with rotation, some quantum advantage can be gained show that our algorithm can work for a diversity of states.

\section{Outlook and Conclusions}

We have presented a machine learning algorithm to conduct quantum teleportation, and demonstrated with a collective spin model that it can be applied to a wide variety of states. This includes single qubits to large particle numbers, systems with prior distributions and unequal dimensions, and coherent states and Dicke states. In all cases, we compared with classical teleportation counterparts where entanglement is not used, and showed  significant regimes of quantum advantage.

The essential feature of our model is its generality, wherein diverse states and scenarios can be addressed. By adjusting the conditions to match any established algorithm tailored for some specific scenario, comparable fidelities can be achieved as we demonstrated with single qubit transfer. Even in such cases, the alogrithm provides the choice of reaching target fidelities  lower than optimal but computable faster. The alogrithm is flexible, so that it can be refined in several ways. Fidelity of teleportation can be improved by using larger sets of search parameters; for example, the bases for measurement outcomes can be optimized as well. That would typically require more computational resources, which in can turn can be optimized as well by utilizing innovations in the arena of artificial intelligence such as deep Q-learning \cite{mnih_human-level_2015}.

\begin{acknowledgments} This work was supported by the Czech Science
Foundation Grant No. 20-27994S for T. Opatrn\'y and by the NSF under Grant No. PHY-2309025 for Kunal K. Das.  \end{acknowledgments}

\appendix

\section{Calculating Variance Boundaries} \label{App:Var}

We start from the quadratic Hamiltonian as described in Eq.(1) of the paper \cite{Opatrny-Das} with $\chi = 1$,
\begin{eqnarray}
    \hat{H}_Q = \left(\hat{J}_{x1}+\hat{J}_{x2}\right)^2 + \left(\hat{J}_{y1}+\hat{J}_{y2}\right)^2
\end{eqnarray}
and a coherent state at a pole of the Bloch sphere
\begin{eqnarray}
    \langle \hat{J}_{xi} \rangle = \langle \hat{J}_{yi} \rangle =\langle \{\hat{J}_{xi},\hat{J}_{yi} \} \rangle = 0,
\end{eqnarray}
with $i=1,2$. These mean values remain zero upon time evolution by $\hat{H}_Q$, as there is no preferred direction. During the evolution, the observables of species $1$ become correlated with those of species $2$. Due to the symmetry, without loss of generality, we pick $\hat{J}_{x1}$ and study its correlations and anti-correlations with a rotated component $\hat{J}_{x2}\cos \phi + \hat{J}_{y2}\sin \phi$. Assume at a given time that the relevant variances are given by
\begin{eqnarray}
    \langle \hat{J}_{xi}^2 \rangle = \langle \hat{J}_{yi}^2 \rangle \equiv \sigma_i^2, \nonumber \\
    \left\langle \left(\hat{J}_{x1} \pm (\hat{J}_{x2}\cos \phi + \hat{J}_{y2}\sin \phi) \right)^2 \right\rangle \equiv \sigma_\pm^2.
\end{eqnarray}
Combining these equations,
\begin{eqnarray}
    \sigma_1^2 + \sigma_2^2 \pm 2\langle\hat{J}_{x1}\hat{J}_{x2}\rangle\cos\phi \pm 2\langle\hat{J}_{x1}\hat{J}_{y2}\rangle\sin\phi = \sigma_\pm^2.
\end{eqnarray}
Rearranging, we find
\begin{eqnarray}
    2(\sigma_1^2 + \sigma_2^2) = \sigma_+^2 + \sigma_-^2, \nonumber \\
    4\left( \langle\hat{J}_{x1}\hat{J}_{x2}\rangle\cos\phi + \langle\hat{J}_{x1}\hat{J}_{y2}\rangle\sin\phi \right) = \sigma_+^2 - \sigma_-^2.
\end{eqnarray}
Optimizing $\phi$ such that the difference $\sigma_+^2 - \sigma_-^2$ is maximized,
\begin{eqnarray}
    &&\tan\phi = \frac{\langle\hat{J}_{x1}\hat{J}_{y2}\rangle}{\langle\hat{J}_{x1}\hat{J}_{x2}\rangle}, \nonumber \\
    &&\sigma_+^2 - \sigma_-^2 = 4 \sqrt{\langle\hat{J}_{x1}\hat{J}_{x2}\rangle^2 + \langle\hat{J}_{x1}\hat{J}_{y2}\rangle^2}.
\end{eqnarray}
This gives the upper and lower boundaries for the variances:
\begin{eqnarray}
    \sigma_\pm^2 = \sigma_1^2 + \sigma_2^2 \pm 2\sqrt{\langle\hat{J}_{x1}\hat{J}_{x2}\rangle^2 + \langle\hat{J}_{x1}\hat{J}_{y2}\rangle^2}.
\end{eqnarray}

\section{Numerical Simulations of Teleportation Algorithm} \label{App:Details}

Here, we provide exactly the steps used to replicate the simulations and data presented in our paper. The information provided in this section includes various initial conditions, times of evolution, state sampling, and minimization functions. The steps parallel those as provided in the algorithm outline.

1. Both systems A and B are started in the lowest-energy eigenstate of $\hat{H}_L$ with $\kappa_{zA}=-\kappa_{zB}=-1$, and all other coefficients set to zero. The exact time of evolution is $t=\pi/4$ for single particle simulations, chosen because it coincided with the time of evolution to maximum entropy. For multi-particle simulations, $t=0.094$ is used to entangle systems A and B. This was chosen as a tradeoff between minimizing measurement variable variance and maximizing entropy, as well as yielding a high fidelity when including a prior distribution. Other such times could be found; this value is not particularly unique, but suited the simulations presented in the paper. For entangling A and C, the time of evolution is different depending on if SU(1,1) or SU(2) interactions are chosen. For SU(1,1), the interaction strength is switched from positive to negative, and the same $t=0.094$ is used. For SU(2) and NOT state teleportation, the interaction strength remains the same, and $t=\frac{\pi}{4N_A}$, where $N_A$ is the number of particles in system A. This maximizes the system entropy for SU(2) interactions.

2. The states C are sampled from a uniform distribution on the Bloch sphere. Two hundred values of $\theta$ are sampled linearly on the interval $[0,\pi]$. For each value of $\theta$, the values of $\phi$ are taken on the interval $[0,2\pi]$, once again using linear sampling. The number of samples of $\phi$ depend on the value of $\theta$:
\bn
    n_{\phi} = 2 \left \lfloor{n_{\theta}\sin\theta}\right \rfloor,
\en
where $n_{\theta} = 200$ for our simulations, and $\lfloor\cdots\rfloor$ is the floor function.

3. The measurements on A and C are done by projecting an ABC state vector onto the given measurements, resulting in a state B vector, which is then normalized.

4. The angles are found using \textit{fmincon}, an inbuilt function provided by MATLAB. The function to minimize is
\begin{eqnarray}
    f(\mathbf{x}) = -\langle \psi_C | \hat{U}(\mathbf{x}) | \psi_B \rangle,
\end{eqnarray}
where $\mathbf{x}$ is a vector with elements given by the angles of rotation. The vector is either two or three elements long as defined by the type of unitary operator being searched for.

Search ranges for all the variables were on the interval $[-\pi,\pi]$. The initial conditions in the multi-particle simulations are uniformly zero for each element of $\mathbf{x}$. In the $N=1$ case, there is an analytical answer. The quadratic Hamiltonian evolves states A and B into the state
\begin{eqnarray}
    |\Phi_{AB}\rangle = \frac{i}{\sqrt{2}}\left(-|\Phi_2\rangle + |\Phi_3\rangle \right),
\end{eqnarray}
where $|\Phi_i\rangle$ denote the four Bell states with $i\in [1,4]$. From this, the exact angles used in the unitary operator to give perfect teleportation fidelity are straightforward to find:
\begin{eqnarray}
    \alpha &=& [\pi/8, 0, 0, -\pi/8] \nonumber \\
    \beta &=& [\pi/2, 0, 0, \pi/2] \nonumber \\
    \gamma &=& [-\pi/8, \pi/4, -\pi/4, \pi/8],
\end{eqnarray}
where the $i^{th}$ index corresponds to measuring $|\Phi_i\rangle$ on the AC subsystem. The data marked by (I) in Fig. 1 use initial conditions given by adding $\epsilon = 0.01$ to each true value. Moving the initial conditions closer to the analytical values makes the teleportation fidelity move closer to unity.

For the data marked by (II) in Fig. 1, the initial conditions, up to two decimal places, are given by
\begin{eqnarray}
    \alpha_0 &=& [1.61, 0.54, 0.81, 1.18] \nonumber \\
    \beta_0 &=& [1.34, -0.72, 0, 3.14] \nonumber \\
    \gamma_0 &=& [-2.16, -2.35, -0.24, -0.67].
\end{eqnarray}

In the case of only using two angles for the unitary operator, data marked by (III) in Fig. 1, the initial values of $\theta_x$ and $\theta_y$ are $\pi/4$ and $\pi/2$, respectively, regardless of which Bell basis vector is measured.

The minimization algorithm is set to Sequential Quadratic Programming (SQP). All other parameters and options for $fmincon$ are set to their default as determined by MATLAB 2023a.

5-6. The procedure here is conducted exactly as stated in the main text.

7. Averaging is conducted by effectively finding the center of mass of solutions, and taking the angular component. In mathematical notation, we solve for the angle $\alpha_n$ satisfying
\bn
    Ae^{i\alpha_n} = \sum_m e^{i\alpha_{nm}}p(\psi_m).
\en

%\bibliography{ML_teleport_refs}


\end{document}